\documentclass[a4paper,fleqn]{cas-sc}
\usepackage{natbib}

\usepackage{babel}
\usepackage{hyperref}
\def\tsc#1{\csdef{#1}{\textsc{\lowercase{#1}}\xspace}}
\tsc{WGM}
\tsc{QE}

\begin{document} 
\let\WriteBookmarks\relax
\def\floatpagepagefraction{1}
\def\textpagefraction{.001}

\shorttitle{}

\shortauthors{Y. Fettach et~al.}

\title [mode = title]{ 
Skill Demand Forecasting Using Temporal Knowledge Graph Embeddings
}
%
\author[1]{Yousra Fettach}

\cormark[1]

\cortext[cor1]{Corresponding author}

\fnmark[1]
\fntext[fn1]{These authors contributed equally to this work. This work has been done at the International University of Rabat.}

\ead{yousra.fettach@uir.ac.ma}


\credit{Conceptualization, Software, Formal analysis, Investigation, Methodology, Writing - original draft}

\affiliation[1]{organization={Ghent University, IDLab, Department of Electronics and Information Systems},
    \city={Ghent},
   country={Belgium}}
   
\affiliation[2]{organization={COLCOM, University Mohammed VI Polytechnic},
    \city={Rabat},
   country={Morocco}}

\author[2]{Adil Bahaj}
\fnmark[1]

\ead{adil.bahaj@uir.ac.ma}
\credit{Software, Investigation, Methodology, Writing - original draft}

\author[2, 3]{Mounir Ghogho}
\affiliation[3]{organization={Faculty of Engineering,University of Leeds},
   country={United Kingdom}}

\ead{mounir.ghogho@uir.ac.ma}
\credit{Conceptualization, Formal analysis, Project administration, Supervision, Validation, Writing - review \& editing}

\begin{abstract}
Rapid technological advancements pose a significant threat to a large portion of the global workforce, potentially leaving them behind. In today's economy, there is a stark contrast between the high demand for skilled labour and the limited employment opportunities available to those who are not adequately prepared for the digital economy. To address this critical juncture and gain a deeper and more rapid understanding of labour market dynamics, in this paper, we approach the problem of skill need forecasting as a knowledge graph (KG) completion task, specifically, temporal link prediction. We introduce our novel temporal constructed from online job advertisements. We then train and evaluate different temporal KG embeddings for temporal link prediction. Finally, we present predictions of demand for a selection of skills practiced by workers in the information technology industry. The code and the data are available on our GitHub repository
(\url{https://github.com/team611/JobEd}).
 
\end{abstract}




\maketitle
\section{Introduction}
As the employment landscape transforms due to emerging technologies, students and employees must confront the reality of skill gaps. This discrepancy between the skills possessed by the workforce and those sought after by employers has the potential to result in millions of unfilled positions over the next decade \cite{garcia2022practical} carrying significant economic and social consequences. Precise projections of future labour market conditions can empower current learners to optimize their future productivity and align with employers' requirements. Although the ownership of this gap lies with individuals – students, and employees – numerous entities play a role in nurturing and managing human capital. Educational institutions, government bodies, and businesses can enhance their decision-making regarding human capital when equipped with accurate labour market forecasts. When decision-makers – including students, educational institutions, government bodies, and businesses – rely on recent trends to guide their choices, they implicitly make predictions about future developments \cite{yazdanian2022radar}.




Labour market demand forecasting started as workforce demand forecasting. In this task, researchers try to predict the evolution of labour demand based on previous observations \cite{kwak1977stochastic} \cite{ward1996workforce} \cite{woolard2003forecasting} \cite{chan2006forecasting}. The measure of labour demand in these studies is the number of workers required for a job. However, these approaches suffer from multiple limitations that hinder the generalization and adaptability of the solutions. These approaches try to model the evolution of labour demand which can be limited to a company, sector, region and other contextual information. This stifles the generalization of the data and the approaches to other settings. In addition, focusing on modelling job demand and not skill demand ignores the ever-changing landscape of the job market, which generates new job titles consistently and requires new skills for old job titles. 

Standard forecasting models, like time series and machine learning methods, often treat skills and jobs as isolated entities or rely on simple, direct relationships. KGs represent entities (such as skills and jobs) as nodes and the relationships between them as edges. This structure naturally captures the intricate web of dependencies and interactions, allowing for a more nuanced understanding of how skills are connected to job roles and industry trends \cite{chao2024cross}.

In a practical setting traditional models require the availability of quantitative measures to be used in forecasting for each job and its corresponding skills. This can present a limitation for new jobs, skills or job-skill pairs. However, this is not the case for skill demand forecasting as a temporal link prediction.

To mitigate some of these limitations, later research tried to solve skill demand forecasting by predicting the demand for certain skills in a job title and how they evolve with time \cite{wolf2023generating} \cite{chao2024cross} \cite{vaidya2021data}, \cite{dawson2020predicting} \cite{jaramillo2020word}. These studies however assume the availability of sufficient data that quantifies the demand for a skill of a given job title with time, after which they can only model that particular skill for that particular job title. Consequently, these studies only focus on a set of jobs with established skills and don't generalize to emerging skills. To surpass these limitations we model the semantics of skills and job titles by exploring the high-level relations between them and other attributes that affect the flow of demand in the job market. Compared to the previous settings, this setting allows for the forecasting of skill demands between jobs and skills that haven't been linked historically. To achieve this we model the interactions between jobs, skills and other related concepts as a temporal knowledge graph \cite{cai2024survey} and formulate the forecasting problem as a temporal knowledge graph link prediction problem.


Knowledge Graphs (KGs) play a crucial role in structuring real-world facts, traditionally presenting static snapshots of current knowledge. However, there is a growing interest in Temporal Knowledge Graphs (TKGs) that aim to capture the intricate temporal dynamics of information (\cite{leblay2018deriving},\cite{garcia2018learning}, \cite{lacroix2020tensor}). Recent endeavours in the field of TKGs focus on predicting future missing links, involving a query about a specific time and associated historical facts. For example, a query could be "What team will win the World Cup in 2026?". The goal is to identify the most appropriate entity (such as a football team) from a set of options to complete the missing information. Essentially, TKGs help predict future relationships in evolving knowledge contexts. In this context, temporal KG completion can be useful in the discovery of new relationships between entities. For example, new jobs have limited information available, to mitigate this, KG completion can be used to uncover the new jobs' relation to existing skills or vice versa.

In this work, we explore temporal knowledge graph forecasting to predict the evolution of skill demand in the job market. To achieve this objective we use JobEdKG, a KG introduced in \cite{fettach2024jobedkg}. JobEdKG is extracted from public job postings and online courses. It contains a plethora of entities and relations that link jobs and skills and other related concepts (e.g. courses, companies, sectors...). JobEdKG facts have a temporal validity, which signifies the timestamp of the first job posting where they occur in the source data. The contributions of this work include: 

\begin{itemize}
    \item We formulate the job skill forecasting problem as a temporal link prediction problem. This approach takes advantage of existing high-order relationships between jobs and skills via related concepts (e.g. sector, company, course etc) to predict the evolution of link strength between jobs and skills.
    \item Introducing a novel temporal KG, which contains high-order relationships between jobs, skills and related concepts. To our knowledge, this is the only temporal KG on the subject of job-skill evolution.
    \item Training and evaluating multiple models on the task of temporal link prediction and leveraging these models to infer the temporal evolution of existing and unseen job-skill pairs.
\end{itemize}


The organization of this paper is outlined as follows: Section \ref{relatedwork} provides an examination of the related literature. Section \ref{methods} outlines a brief run-through JobEdKG (\cite{fettach2024jobedkg}) and the description and implementation details of our KG completion model. The experiments we conducted are presented in Section \ref{results} where we also showcase the results of our case study and discuss further implications of our work. Finally, we conclude the paper in Section \ref{conclusion}.

\section{Background and Related work}
\label{relatedwork}



The recent emphasis on skills has become possible due to advancements in natural language processing (NLP) and the digitalization of texts related to the labour market. Job advertisements offer valuable insights into labour demand, while resumes and social media profiles provide information about the skills possessed by different types of workers. In the next few subsections, we explore skill demand forecasting, the use of knowledge graphs to analyze the job market and temporal KG embeddings in comparison with time series forecasting.

\subsection{Skill Demand Forecasting}

Research on predicting skill demand and supply can be broadly classified into two categories: expert-based and learning-based methods. Expert-based approaches analyze skill shifts using survey data, often providing general forecasting results based on expert domain knowledge. Learning-based methods, on the other hand, offer more detailed predictions by leveraging machine learning techniques to identify patterns and implications. Expert-based studies started with studies such as \cite{ward1996workforce} \cite{woolard2003forecasting} \cite{chan2006forecasting} that focused on workforce demand forecasting which faced many limitations because of the lack of generalization in these methods. 

Learning-based methods focused on skill demand forecasting. Studies like \cite{vaidya2021data} used the SVM model to predict the skills based on profession. Others like \cite{dawson2020predicting} used the  XGBoost to predict yearly skills shortages in labour markets. Studies such as \cite{jaramillo2020word} built a spatial representation of the job market that allows for tracking change in skills' demand, however,  this representation didn't incorporate any temporal information. Similarly, \cite{liu2021learning} used multiple neural networks to match jobs to skills. Time series analysis approaches were also utilized in this direction of research, where \cite{garcia2022practical} used time series for practical skills demand forecasting and \cite{sibarani2020scodis} presented a methodology for the discovery of in-demand skillsets based on the observation of skills clusters in a time series.


\subsection{Knowledge graphs in analyzing the job market}

Studies such as (\cite{luo2019learning}) have tackled the issue of employment representation in KGs. In order to capture the similarity and ordering relationships between job titles and business pairings, the authors created a representation of employment roles. The Skills \& Occupation KG developed in (\cite{de2021job}) is the most recent study in this field of job advice. This KG was created by incorporating external job ads into the taxonomies of skills and occupations that already existed. Another study created a KG to represent the artificial intelligence job market and the accompanying necessary abilities (\cite{jia2018representation}). The KG made it easier to investigate the connections between various AI-related key competencies. JobEdKG is another important piece of literature \cite{fettach2024jobedkg} where the authors built a KG based on job postings and Massive Open Online Courses (MOOCs) to recommend online courses and predict in-demand skills based on uncertain KG embeddings. 
\subsection{Temporal KGs embedding}
KG embeddings aim to generate low-dimensional representations of entities and relationships \cite{wang2017knowledge}. These embeddings rely on different components such as the embedding space and the scoring function that measures the plausibility of factual triples. 

Recent methods have taken into consideration the temporal information for more accurate embeddings and better link prediction. Inspired by the taxonomy in \cite{han2021time}, we divide the literature on TKG embedding into three categories: timestamp embedding, time-dependent entity embedding and deep temporal representation. 
	
Timestamp embedding (TE) models learn representation for each discrete timestamp. TTransE \cite{leblay2018deriving} is an example, which extends TransE by adding one more embedding function mapping timestamps to hidden representations. This kind of embedding has several limitations since it leads to more synthetic relations than necessary in the KG. Other models such as HyTE \cite{dasgupta2018hyte} explored projections. HyTE incorporates time in the entity-relation space by associating each timestamp with a corresponding hyperplane.
	
Time-dependent entity embedding (TEE) models are built on taking time and the entity as inputs to the entity embedding function and providing a hidden representation, that incorporates time, as an output. Such models include DE-SimplE \cite{goel2020diachronic}, which was inspired by the success of the diachronic word embeddings, and AtiSe \cite{xu2020temporal}, which incorporates temporal information into entity/relation representations by using additive time series decomposition. Other models such as TeRo \cite{xu2020tero}, 3DRTE \cite{wang20203drte} and RotateQVs \cite{chen2022rotateqvs} used rotations to incorporate time in the entity-relation embedding. TeRo applies this in the complex space whereas 3DRTE and RotatQVS do the same in the quaternion space. 
	
Using deep temporal representation (DTR) learning to integrate time was also an important direction that certain studies have taken.  \cite{garcia2018learning} utilized recurrent neural networks (RNN) to incorporate temporal information into the relation representations of DistMult and TransE. By feeding time characters and the static relation embedding to an LSTM, the model learns a temporal embedding for each relation.  
	

\subsection{Time series forecasting vs. TKG link forecasting}
There is a distinction between time series forecasting and TKG forecasting. Time series forecasting entails the prediction of future measures from past measures \cite{chen2024job}, while TKG forecasting is the task of learning the temporal dynamics and how they affect the semantics of TKG entities and relations \cite{du2023imf}. Specifically, in time series forecasting time is used as an index to the observations of a measure to order them sequentially, while in TKG forecasting it is considered a feature and its influence on the semantics of the TKG is modelled directly. From another perspective, TKG forecasting is a self-supervised task, while time series forecasting is a supervised task, meaning that time series forecasting requires the existence of high-quality supervision data to work, which can be a limiting factor in the case of job/skill demand forecasting since they are significantly sparse \cite{chen2024job}. However, by directly modelling the temporal dynamics of entities and relations, TKG forecasting surpasses the inherent sequential limitations of forecasting approaches and the need for supervised measures. In addition, since time series forecasting and TKG forecasting solve totally different tasks under different assumptions, comparing these approaches is pointless and can't be done practically.

\section{Methodology}
\label{methods}
In this section, we briefly go through the pipeline of JobEdKG \cite{fettach2024jobedkg} construction before we explain the temporal KG embedding methods we experimented with.

\subsection{Data: JobEdKG}
We constructed JobEdKG \cite{fettach2024jobedkg} using online job postings and courses. We compiled job postings from the Moroccan website Rekrute.com \footnote{https://www.rekrute.com/}, spanning from November 2005 to September 2022. Each job posting in our dataset includes various details such as the job title, cognitive (hard) and non-cognitive (soft) skills required, the hiring company, the economic sector associated with the job, the job's location, educational prerequisites, required experience, contract type, job function, posting date, and expiration date. To ensure data accuracy and prevent duplicate entries, we used unique identifiers from recruiter companies found on the website. We retained only those job postings that offered positions within Moroccan cities. The final dataset consists of 99,676 job postings, encompassing both English and French content. Additionally, we collected Massive Open Online Courses (MOOCs) from Coursera using the website's API. The resulting MOOC dataset comprises a total of 10,180 courses, featuring course names, descriptions, specializations, and instructor information.

In order to extract hard skills from the different job postings and course descriptions we opted for a rule-based approach that uses existing databases of hard skills and extracts skills using the Spacy library rule mining module \footnote{https://spacy.io/usage/rule-based-matching}. We extracted hard skills in both French and English. For the hard skills mentioned in English, we used an existing set of rules from Jobzilla \footnote{https://github.com/pandmi/jobzilla\_ai}. For the hard skills mentioned in French, we constructed our own set of rules based on the ESCO \footnote{https://esco.ec.europa.eu} and ROME \footnote{https://www.pole-emploi.fr/employeur/vos-recrutements/le-rome-et-les-fiches-metiers.html} hard skills datasets.

For the rule-building process from raw skills data, for each hard skill, we applied an exact word-building process and a POS (Part-of-speech) tagging-based process.
After extracting the hard skills using the rules, we identify the incompleteness of a significant number of them, so we apply sentence filtering using the TextAttack model \footnote{https://huggingface.co/textattack/roberta-base-CoLA}. Seeing that this model is trained
on English data, we translated the extracted skills using two translators: Argos translator \footnote{https://github.com/argosopentech/argos-translate} and a transformer-based translator \footnote{https://huggingface.co/Helsinki-NLP/opus-mt-fr-en}. After translating the skills, we use the pre-trained TextAttack model to identify if the sentence (skill in our case) is complete or not in addition to quantifying the degree of their completeness with an acceptability score \cite{fettach2024jobedkg}.
Relations are extracted based on the co-occurrence of different types of entities in the same job offer or the same course. For example, if a job offer describes a job title belonging to a certain sector, then that job title and sector are linked. The outcomes of skill entity recognition were combined with relation extraction to create a dataset that includes hard skills, soft skills, economic sector, recruiter company, function,
and courses for each job title.

JobEdKG was constructed as an uncertain KG, meaning each link between the nodes has a score of uncertainty but for this work, we don't consider these scores and we focus on the timestamps for each link that indicates the validity of the fact at a certain time point so for example if the job "manager" required the skill "Excel" in the timestamp "11/11/2006" that link would be present in the KG starting that time. To create some sort of differentiation between the temporal JobEDKG and JobEDKG. We refer to the knowledge graph in this paper by T-JobEDKG.

Since the source data is well structured we didn't need to apply advanced entity resolution techniques. For job titles, we used fuzzy matching of job titles, such that if two job title names Levenshtein distance \cite{navarro2001guided} is below a threshold they are considered similar. For all the other entities we used exact matching. We were able to use exact matching in entities such as courses, companies, and soft skills because they are standardised from the source websites. In the case of hard skills, the extraction process is rule-based and it maps the extracted hard skills to a finite set of hard skills from the ESCO and ROME datasets. Consequently, we were able to use exact matching for entity resolution.

\begin{figure*}[ht]
\centering
  \includegraphics[width=15cm]{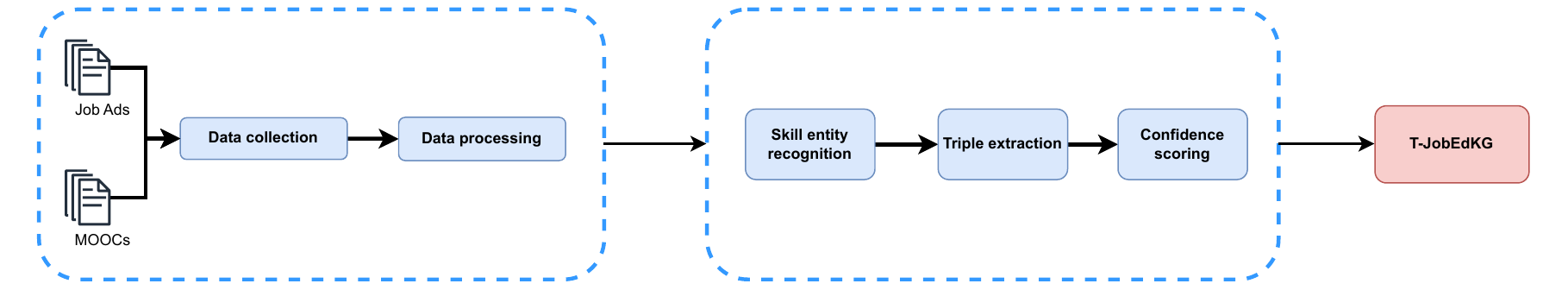}
  \caption{Overview of the methodology for building T-JobEdKG}
  \label{tjobedkg}
\end{figure*}

\begin{figure*}[ht]
\centering
  \includegraphics[width=15cm]{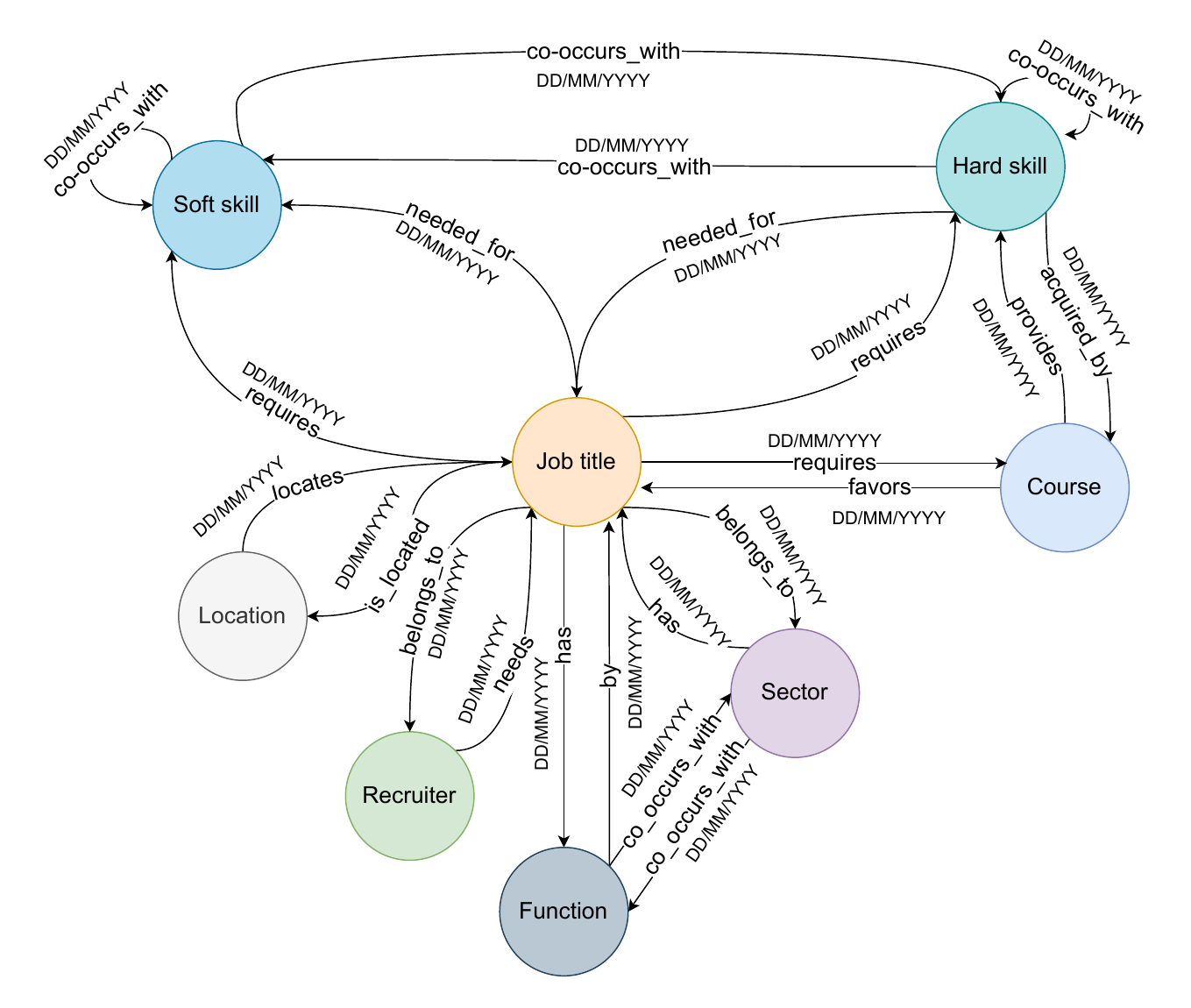}
  \caption{Metamodel of T-JobEdKG. DD/MM/YYYY stands for the timestamps associated with links (day, month, year)}
  \label{metamodel}
\end{figure*}

The knowledge graph (KG) resulting from the process outlined in Figure \ref{tjobedkg} and discussed in the previous section comprises a total of 55718 nodes. The number of nodes for each entity type is showcased in Table \ref{node_types}. These nodes are interconnected by a network of 1296374 relations across 12 distinct types. The metamodel for our KG is in Figure \ref{metamodel}.


\begin{table}[]
\centering
\resizebox{0.3\textwidth}{!}{
\begin{tabular}{|l|l|} 
\hline
Entity types & Number of nodes \\ \hline
Job titles & 9736 \\ \hline
Hard skills & 32824 \\ \hline
Soft skills & 21 \\ \hline
Courses & 10180 \\ \hline
Locations & 35 \\ \hline
Recruiters & 2714 \\ \hline
Sectors & 133 \\ \hline
Functions & 75 \\ \hline
\end{tabular}%
}
\caption{Entity types and number of nodes of T-JobEdKG}
\label{node_types}
\end{table}

The location entities correspond to the cities where companies that posted job offers are situated. Recruiters' entities denote the names of the companies that posted the job offers, while sector nodes represent the economic sectors featured on the Rekrute website. The relationships established between job titles and entities of the mentioned types convey semantic significance. These relationships fall under the 'belongs\_to' relationship type, connecting job titles as the primary entities to the companies as the secondary entities. This signifies the job's association with a specific company.

The 'requires' relation links jobs as primary entities to the skills as secondary entities, indicating the skills necessary for a particular job. The 'located' relation associates jobs as primary entities with locations as secondary entities, while the 'has' relation connects job titles as primary entities to their respective functions.

Other relation types link skills to one another, such as 'co\_occurs\_with,' which establishes connections between skill entities based on their co-occurrence in the same job offer, or to different types of entities, like 'acquired\_by,' which connects skills as primary entities and courses as secondary entities. The 'provides' relation associates MOOCs as primary entities with skills as secondary entities to convey the skills offered by a specific course, while the 'favors' relation links MOOCs as primary entities with job titles as secondary entities, indicating preferences or affinities between them.

\subsection{Temporal KG Link Prediction}
Temporal KG forecasting predicts future links based on past facts. Practically, it implies predicting the tail $t$ based on a question $(h, r_{q}, ?, \tau_{q})$ and the previous facts $\mathcal{G}_{\tau_{q}}$, where $r_{q}$, $\tau_{q}$ denote the relation and timestamp of the question. Temporal KG forecasting involves ranking all entities of the specific timestamp and obtaining the most likely links given the timestamp. We present the preliminaries for this as follows:
 \paragraph{\textbf{Temporal KG.}} Suppose that $\mathcal{E}$, $\mathcal{R}$, and $\mathcal{T}$ represent the entity set, relation set, and timestamp set, respectively. The temporal KG is a collection of quadruples, which can be expressed as
    \begin{equation}
        \mathcal{K} = \{(h, r, t, \tau), h,t \in \mathcal{E}, r \in \mathcal{R}, \tau\in \mathcal{T}\} 
    \end{equation}
    $(h, r, t, \tau)$ denotes a quadruple; $h$ and $t$ represent the head and tail, respectively. $r$ represents the relation, and $\tau$ represents the time that the quadruple occurs. 
\paragraph{\textbf{The task of temporal KG forecasting.}} Suppose that facts happening before the selected time $\tau_{k}$ can be expressed as
\begin{equation}
    \mathcal{G}_{\tau_{k}} = \{(h_{m}, r, t_{n}, \tau_{i}) \in \mathcal{K}|\tau_{i} < \tau_{k}\}
\end{equation}

\begin{figure}
    \centering
    \includegraphics[scale = 0.39]{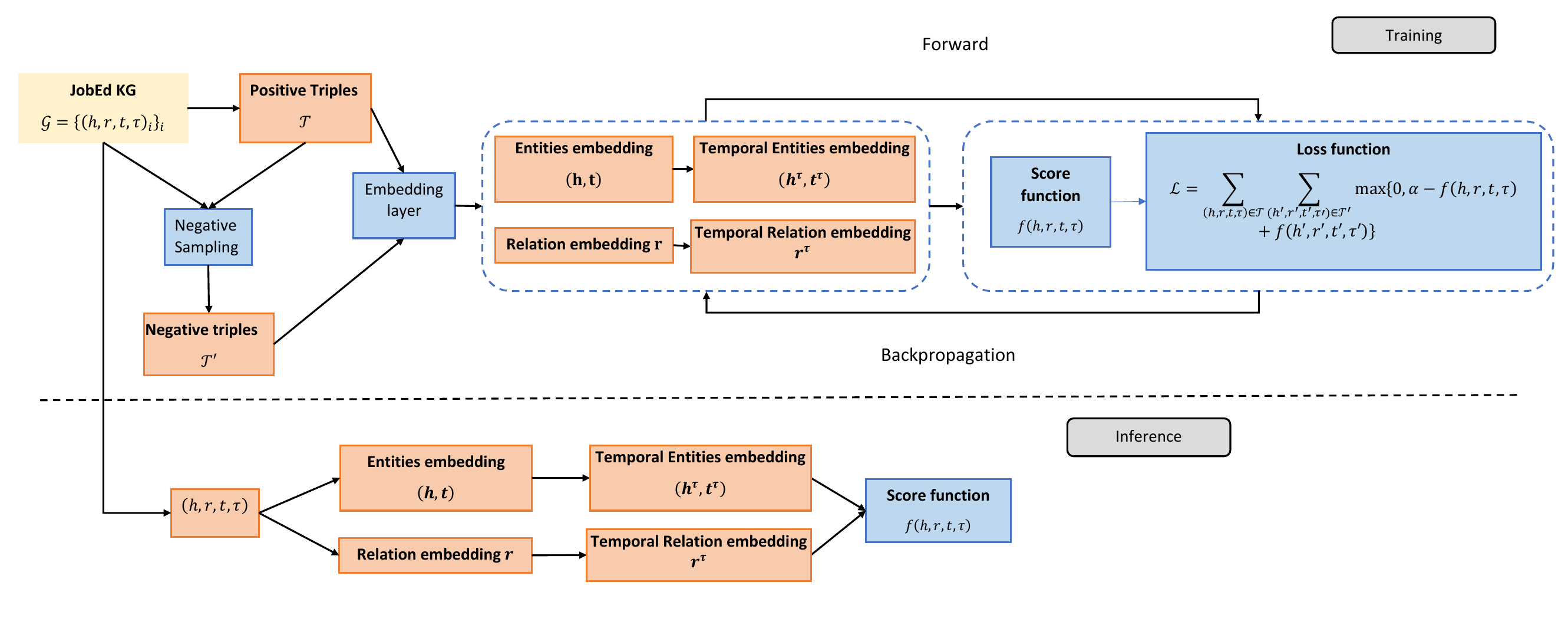}
    \caption{Training and inference on temporal KGs. The model is trained on triples with timestamps. It learns temporal embeddings to rank correct triples higher than incorrect ones. After training, it can predict missing or future relations.}
    \label{fig:tempo_kge}
\end{figure}

Temporal KG link prediction is the task of predicting missing links from existing ones. To achieve that we embed entities and relations to low dimensional representation space and transform those structural representations using temporal information to obtain temporal embeddings of entities or/and relations. These embeddings are then fed to a scoring function to obtain the likelihood of the existence of a relation between the two entities given the time of observation. Figure \ref{fig:tempo_kge} summarizes the temporal KG link prediction training and inference process. The process starts with training, where the model is provided with triples in the form of (head, relation, tail, timestamp). During training, the model learns temporal embeddings for entities and relations by optimizing a scoring function that sampled via negative sampling, which ranks the correct triples higher than incorrect (negative) triples. Temporal aspects are integrated through methods such as time-aware embeddings (DE-TransE, DE-DistMult) or recurrent models (TA-TransE, TA-DistMult) that account for changes in relationships over time. Once trained, the model undergoes inference on unseen data. Given a query (e.g., predicting a future relation), the model generates predictions by ranking possible tails for a (head, relation) pair or predicting missing relations. In what follows the detail the approaches used to embed the temporal KG.
\subsubsection{Embedding layer}
The objective of the embedding layer is to represent entities and relations in a low-dimensional space that preserves the semantic and structural properties of the KG. This work explores multiple approaches that incorporate temporal information in entity and relation embeddings. \cite{goel2020diachronic} follows TEEs, by embedding entities and relations in $ \mathbb{R}^{d} $. Entities are embedded by having a portion being structurally dependent and a portion of the embedding is time dependent as follows: the embedding $ \mathbf{z}^{\tau}_{e} $ of an entity $ e $ is defined as \begin{equation*}\label{key}
        \mathbf{z}_{e}^{\tau}[n]=\begin{cases}
            \mathbf{a}_{e}[n]\sigma(\mathbf{w}_{e}[n]\tau + \mathbf{b}_{e}[n])& \text{ if } 1 \le n \le \gamma d\\
            \mathbf{a}_{e}[n]& \text{ if } \gamma d < n \le d
        \end{cases}
    \end{equation*}
    where $\mathbf{a}_{e}\in\mathbb{R}^{d}$, $\mathbf{w}_{e}, \mathbf{b}_{e}\in\mathbb{R}^{\gamma d}$ are entity specific embeddings, $\sigma$ is an activation function, $ 0 \le \gamma \le 1 $ is a hyper-parameter that controls the percentage of temporal features.
    \cite{garcia2018learning} follows DIR by incorporating temporal information in the relation embedding. This is done by extending a relation $ r $ to a sequence $ r_{seq} $, which contains $ r $ along a sequence that represents time $\tau$ as follows: $ r_{seq}=(r,T_{1y},T_{2y},T_{3y},T_{4y}, T_{1m}, T_{2m},T_{1d},T_{2d}) $, where the suffixes $ y $, $ m $, and $ d $ indicate whether the digit corresponds to the year, month, or day information. Let $ \mathbf{r}_{seq} \in \mathbb{R}^{d}$ be the embedding of $ r_{seq} $. In order to obtain $ \mathbf{r}_{seq} $ each token of the sequence $ r_{seq} $ is first mapped to its corresponding $ d $-dimensional embedding and the resulting sequence of embeddings used as input to the LSTM. $ \mathbf{r}_{seq} $ is presented by the last hidden state of the LSTM. More recently, \cite{xu2020tero} proposed Tero, an approach that uses the complex embedding space. Complex space approaches have achieved state-of-the-art performance in TKG embedding. In Tero each element of a triple $(h,r,t,\tau)$ is represented by a complex vector $ \mathbf{h},\mathbf{r},\mathbf{t},\mathbf{\tau} \in \mathbb{C}^{d} $. To obtain time-aware head and tail representations Tero uses the Hamiltonian product:
    \begin{equation}
        \mathbf{z}^{\tau}_{h}=\mathbf{h}\circ\mathbf{\tau}\textbf{ ; }
        \mathbf{z}^{\tau}_{t}=\mathbf{t}\circ\mathbf{\tau}
    \end{equation}

\subsubsection{Scoring function}

The objective of a score function is to quantify the likelihood of two entities being linked via a relation. There are multiple classes of scoring functions in the literature that achieve that: mainly we have distance-based and semantic scoring functions. Distance-based scoring functions evaluate the plausibility of a triple by measuring the distance between transformed entity embeddings in the vector space. The goal is to make the embeddings of related entities close to each other after applying a relation transformation. Examples of distance-based models include TransE and its variants. Semantic scoring functions evaluate the plausibility of a triple by using interactions or similarities between the embeddings of the entities and the relation. These functions often involve multiplicative or additive interactions among the embeddings. Examples include DistMult, ComplEx, and RESCAL. In our case, we choose a scoring function for each class: TransE is distance-based while DistMult is semantic.
    
    The previously presented temporal embeddings of relations and entities are used in different scoring functions to calculate the likelihood of a fact. These scoring functions are an extension of the TransE \cite{bordes2013translating} and the DistMult \cite{yang2015embedding} scoring function, which can be described as follows for a fact $(h,r,t)$ $f(h,r,t)=||\mathbf{h}+\mathbf{r}-\mathbf{t}||_{2}^{2}$ and $f(h,r,t)=\mathbf{h}^{\top}\cdot(\mathbf{r}\odot\mathbf{t})$ respectively where $\mathbf{h},\mathbf{r},\mathbf{t}\in\mathbb{R}^{d}$ are embeddings of $h, r$, and $t$ respectively. We experimented with the following scoring functions: let $\mathbf{z}_{e}^{\tau}$ be the temporal embedding of an entity $e$ at time $\tau$, $\mathbf{r}$ is the static embedding of relation $r$, $\mathbf{r}_{\tau}$ is the dynamic embedding of relation $r$, $\mathbf{h}$ and $\mathbf{t}$ are static embeddings of $h$ and $t$ respectively.
    \begin{itemize}
        \item DE-TransE \cite{goel2020diachronic}:
        \begin{equation}
            f(h,r,t,\tau)=||\mathbf{z}_{h}^{\tau}+\mathbf{r}-\mathbf{z}_{t}^{\tau}||_{2}^{2}
        \end{equation}
        \item DE-DistMult \cite{goel2020diachronic}:
        \begin{equation}
            f(h,r,t,\tau)=(\mathbf{z}_{h}^{\tau})^{\top}\cdot(\mathbf{r}\odot \mathbf{z}_{t}^{\tau})
        \end{equation}
        \item TA-TransE \cite{garcia2018learning}:
        \begin{equation}
            f(h,r,t,\tau)=||\mathbf{h}+\mathbf{r}_{seq}-\mathbf{t}||_{2}^{2}
        \end{equation}
        \item TA-DistMult \cite{garcia2018learning}:
        \begin{equation}
            f(h,r,t,\tau)=(\mathbf{h})^{\top}\cdot(\mathbf{r}_{seq}\odot \mathbf{t})
        \end{equation}
        \item Tero \cite{xu2020tero}:
        \begin{equation}
            f(h,r,t,\tau)=\parallel\mathbf{z}^{\tau}_{h}+\mathbf{r}-\bar{\mathbf{z}}^{\tau}_{t}\parallel
        \end{equation}
    \end{itemize}
    where $\cdot$ is the dot product and $\odot$ is the element-wise product. $\bar{\mathbf{z}}$ is the conjugate of $\mathbf{z}\in\mathbb{C}^{d}$.
\subsubsection{Training objective}
    We trained the previously mentioned models using the Adam optimizer, to minimize the triplet loss:
    \begin{equation}\label{key}
	\mathcal{L}=\sum_{(h,r,t, \tau)\in \mathcal{T}} \sum_{(h',r',t',\tau')\in \text{Neg}(h,r,t,\tau,\mathcal{T}, n)}\max\left\{0,\alpha-f(h,r,t,\tau)+f(h',r',t',\tau)\right\}
	\end{equation}
	with $ \alpha $ is the margin hyperparameter, $ \text{Neg} $ is the negative sampling function, which generates $ n $ negative for each quadruple $ (h,r,t, \tau)\in \mathcal{T} $. In our experiments, we employed a random uniform negative sampling strategy where either the head or the tail of a fact is replaced with a random entity to generate a ransom fact \cite{bordes2013translating}.
 
\subsubsection{Evaluation metrics}
    After training the model using the training objective, the model needs to be adequately evaluated on the temporal link prediction task. In order to do so we use existing metrics from literature \cite{goel2020diachronic}. For a a dynamic fact $ (h,r,t,\tau) $ from the test set $\mathcal{E}_{test}$ we create two queries:  1- $ (h,r,?,\tau) $ and 2- $ (?,r,t,\tau) $. For the first query, we rank all the entities using the scoring function.
	For each quadruple $ (h,r,t,\tau) $, we first take $ (?,r,t,\tau) $ as the query and obtain the filtered rank on the head:
    \begin{equation*}\label{key}
        \text{rank}_{h}=\left|\left\{e\in \mathcal{V}:
        (f(e,r,t,\tau)\ge f(h,r,t,\tau))\wedge((e,r,t,\tau)\notin \mathcal{E}_{train}\cup \mathcal{E}_{val})\right\}\right|+ 1
    \end{equation*}
    Next, we take $ (h,r,?,\tau) $) as the query and obtain the filtered rank on the tail:
    \begin{equation*}\label{key}
        \text{rank}_{t}=\left|\left\{e\in\mathcal{V}:
        (f(h,r,e,\tau)\ge f(h,r,t,\tau))\wedge((h,r,e,\tau)\notin \mathcal{E}_{train}\cup \mathcal{E}_{val})
        \right\}\right|+1
    \end{equation*}
    The following metrics are computed from both the head and tail ranks on all quadruples : (i) Mean reciprocal ranking (MRR):
    \begin{equation*}\label{key}
        MRR=\frac{1}{2|\mathcal{E}_{test}|}\sum_{(h,r,t,\tau)\in\mathcal{E}_{test}}\left(\frac{1}{\text{rank}_{h}}+\frac{1}{\text{rank}_{t}}\right)
    \end{equation*}. (ii) $ H@k $: ratio of ranks no larger than $ k $, i.e.,
    \begin{equation*}\label{key}
        H@k=\frac{1}{2|\mathcal{E}_{test}|}\sum_{(h,r,t,\tau)\in\mathcal{E}_{test}}\left(\mathbb{I}(\text{rank}_{h}\le k) +\mathbb{I}(\text{rank}_{t}\le k)\right)
    \end{equation*}
    where $ \mathbb{I}(a) = 1 $ if $ a $ is true, otherwise $ 0 $. (iii) $ MR $: mean rank:
    \begin{equation*}\label{key}
        MR= \frac{1}{2|\mathcal{E}_{test}|}\sum_{(h,r,t,\tau)\in\mathcal{E}_{test}}\left(\text{rank}_{h}+\text{rank}_{t}\right)
    \end{equation*}
    The larger the $ MRR $ or $ H@k $, the better the embedding. The smaller the $ MR $, the better the embedding.

\section{Results}
\label{results}

\subsection{Implementation details}
    We conducted experiments on four 12GB GPUs. The code that we implemented is based on the TorchKGE library, which we adapted to accommodate temporal knowledge graphs. We were motivated to implement our version since the original implementations for the models were "prohibitively" slow. We made this library publicly available at \url{https://github.com/BahajAdil/TempTorchKGE}. In addition, the source code is available at \url{https://shorturl.at/divxC}, which is part of the larger JobEd project \url{https://github.com/BahajAdil/JobEd}.
    
\subsection{Experimental Setup}
    The aim is to compare the performances of different scoring functions on the T-JobEdKG knowledge graph for the task of temporal link prediction. We conducted a grid search using multiple hyperparameters. For the two models, the following hyper-parameter spaces were used: learning rate: $(0.01, 0.001, 0.0001)$, number of negatives per positive example: $(10,30,40)$, margin:$(1,5, 10,20)$, embedding dimension: $(50,100, 150, 200)$. We fixed the batch size to 2000. For diachronic embeddings, we experimented with two ratios for temporal representation $(0.1,0.2)$.
\subsection{Performance Analysis}
Tables \ref{tab:best_tempo_models} and \ref{tab:best_tempo_hyperparams} show the performance of different models on the task of temporal link prediction and the best-performing hyperparameters. TA models have a compatible performance with each other in the same way DE models have, demonstrating the limited advantage of changing the scoring function on this task. DistMult improved upon the performance of TransE in TA and DE settings. The TA model family outperformed DE model family. The best-performing model in TA-DistMult. We think that various factors explain why TA is better than DE even though DE is the more recent model and performs better than TA in various benchmark datasets. In what follows we give these facts and our interpretation:
\begin{itemize}
    \item Time-aware entity vs. relation embedding: TA incorporates the time element in the relation to have a time-aware relation embedding. DE and Tero incorporate the time element in the head and tail entities to have time-aware head and tail embedding. Consequently, TA models relations change with time while DE and Tero model the changes of entities' semantics with time. Our task is related to job/skill demand and not some task where the meaning/semantics of skills or any other entity change (job transitioning, skill repurposing etc). This explains why TA models performed significantly better in this task.
    \item Model complexity: TA models have a much higher computational complexity than DE models. In DE, time-aware entity embeddings are created as a function of the raw timestamps. At the same time, TA embeds each element of the timestamp into a low dimensional representation space which is then fed to an LSTM (giving it even more weights). Given the known phenomenon of bias-variance trade-off and since TA models contain more variance they tend to be more adept for larger TKGs (more bias), while DE models and Tero are more adequate for sparser and smaller TKGs.
    \item Real embedding space vs. complex embedding space: Another distinctive characteristic along which we can compare the results is the embedding space. The Tero model represents embeddings in the complex space while DE and TA models represent their entities and relations in the real space. From the results, it seems that real embedding space gives better results. This may be attributed to the rotation operations performed in the complex space, which can "overshoot" entity representations in a denser representation space, resulting in entangled representations. This is exactly the case of our KG, which is composed of a larger amount of facts, entities and relations than the standard benchmarks.
\end{itemize}
\begin{table}[h]
            \centering
            \resizebox{0.6\textwidth}{!}{
            \begin{tabular}{lrrrr}
                \toprule
                Model &  Hit@1 &  hit@3 &  hit@10 &  MRR \\
                \midrule
                DE-DistMult &   54.82 &   70.54 &  91.37 &  65.79 \\
                DE-TransE &    48.49 &    58.30 &    71.40 &  56.07\\
                TA-TransE &    61.26 &    79.56 &     90.08 &  71.65 \\
                Tero &48.15&49.53&50.82&49.17\\
                TA-DistMult &   \textbf{61.77} &    \textbf{84.53} &     \textbf{95.08} & \textbf{74.20} \\
                \bottomrule
            \end{tabular}
            }
            \caption{The evaluation metrics of the best performing models (\%).}
            \label{tab:best_tempo_models}
        \end{table}
        \begin{table}
            \centering
            \resizebox{0.7\textwidth}{!}{
            \begin{tabular}{lrrrrr}
                \toprule
                Model &     lr &  emb\_dim &  margin &  n\_neg &  s\_rep\_ratio  \\
                \midrule
                DE-DistMult & 0.0010 &      100 &       1 &     10 &          0.1\\
                  DE-TransE & 0.0001 &      100 &       1 &     40 &          0.1\\
                 TA-DistMult & 0.0010 &       50 &       1 &     10 &          - \\
                Tero&0.001&200&10&100&\\
                   TA-TransE & 0.0010 &       50 &      20 &     30 &          -\\
                \bottomrule
            \end{tabular}
            }
            \caption{The optimal hyperparameters for the different models. lr is the learning rate, emb\_dim is the embedding dimension, n\_neg is the number of negative examples per positive, s\_rep\_ratio is the ratio of temporal features.}
            \label{tab:best_tempo_hyperparams}
        \end{table}
\subsection{Scalability and Convergence Speed}
    To embed our knowledge graph we use a framework that we implemented called TempTorckKGE \footnote{\url{https://github.com/BahajAdil/TempTorchKGE}}, which is an adaptation of TorchKGE \cite{boschin2020torchkge} for temporal knowledge graphs. TempTorckKGE takes advantage of bulk matrix operations instead of single triple operations to calculate the scores, which makes it significantly more efficient than the original implementations of the models that we used and other more recent models (e.g. TeRo \cite{xu2020tero}, Atise \cite{xu2019temporal}).
\subsection{Case study of relation inference}

\begin{figure*}[ht]
\centering
  \includegraphics[width=15cm]{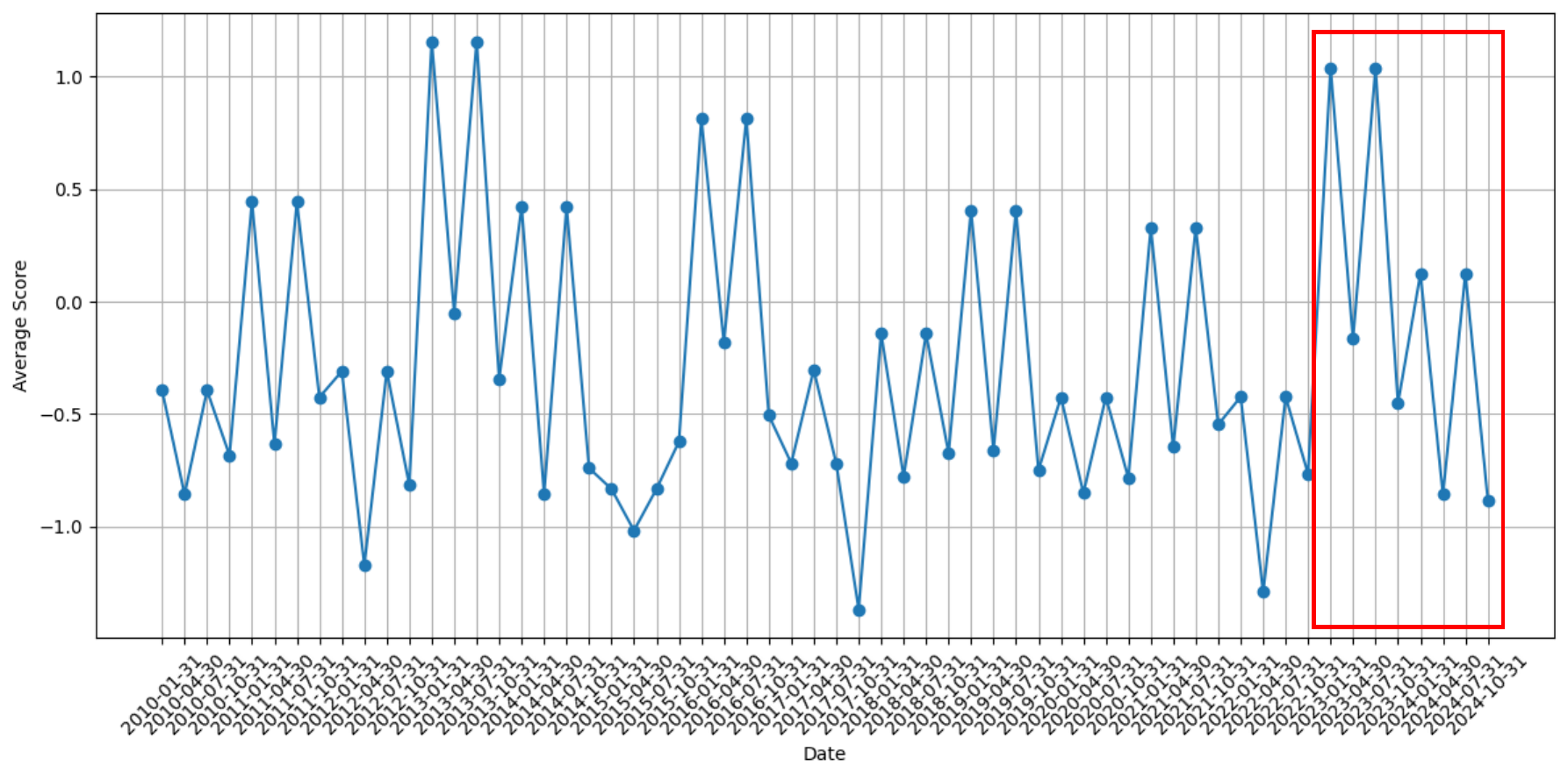}
  \caption{Temporal evolution of the "data analysis" skill for the job "Consulting Engineer in IT Management". The red square showcases the prediction of our model for the time range between 01-01-2023 and 01-01-2025.}
  \label{da_evol}
\end{figure*}

\begin{figure*}[ht]
\centering
  \includegraphics[width=15cm]{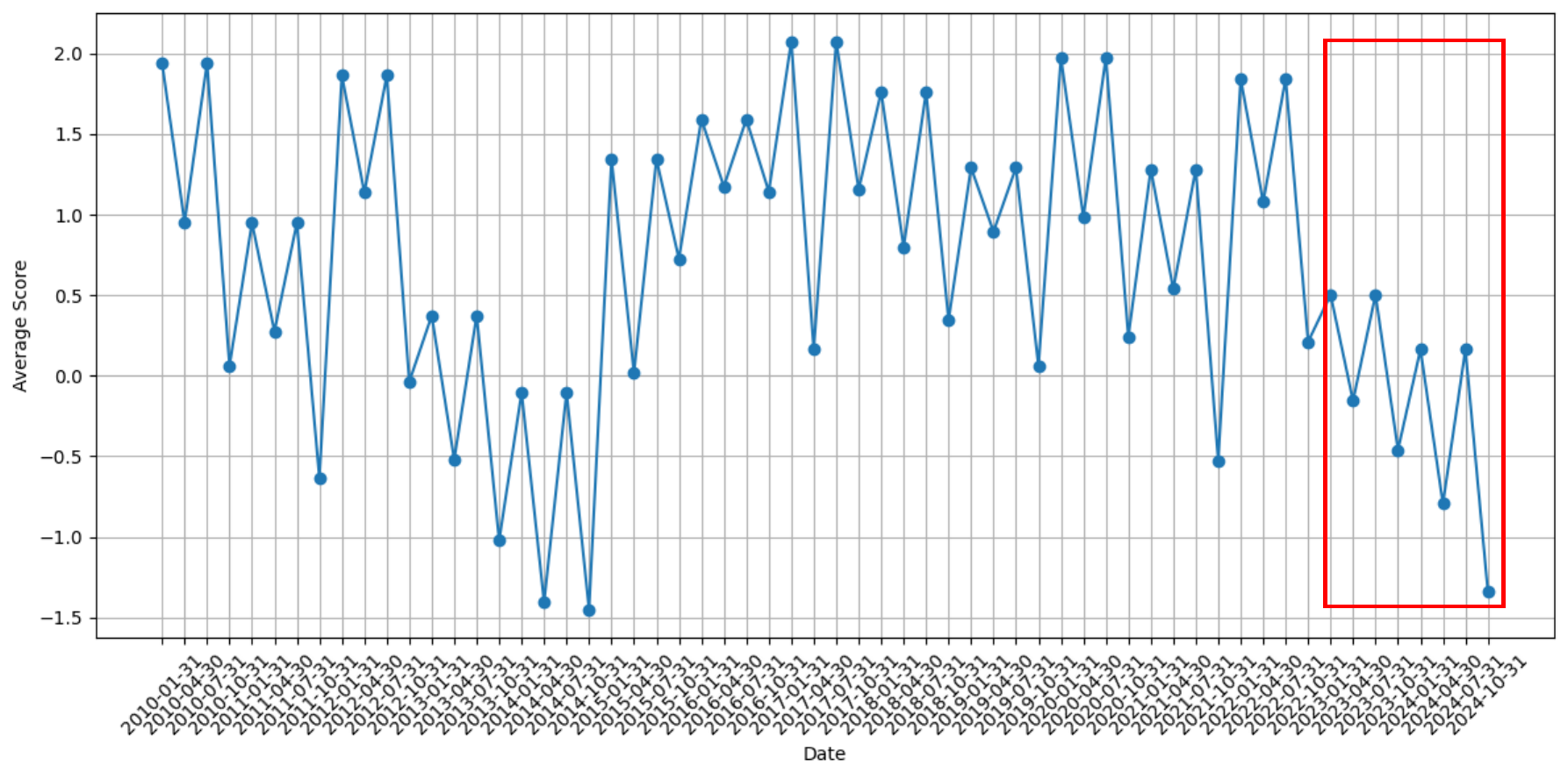}
  \caption{Temporal evolution of the "Java" skill for the job "Consulting Engineer in IT Management". The red square showcases the prediction of our model for the time range between 01-01-2023 and 01-01-2025}
  \label{java_evol}
\end{figure*}

\begin{figure*}[h!]
\centering
  \includegraphics[width=14cm]{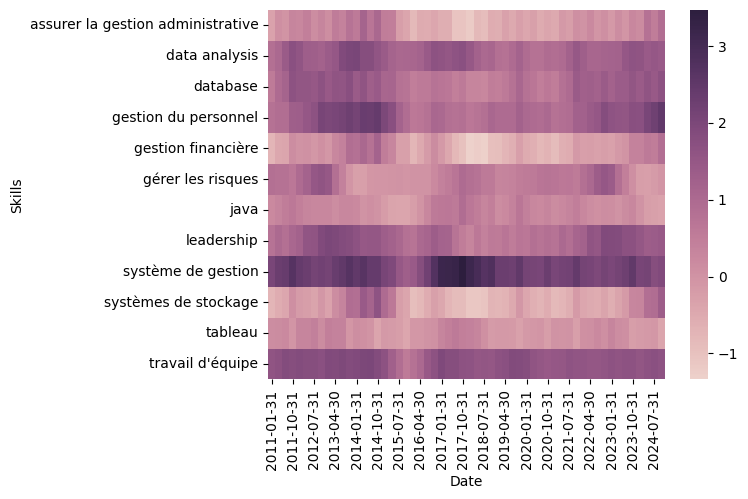}
  \caption{Heatmap showcasing the top skills for the job "Production Leader" and their level of importance through the years.}
  \label{heatmap_1}
\end{figure*}

\begin{figure*}[h!]
\centering
  \includegraphics[width=14cm]{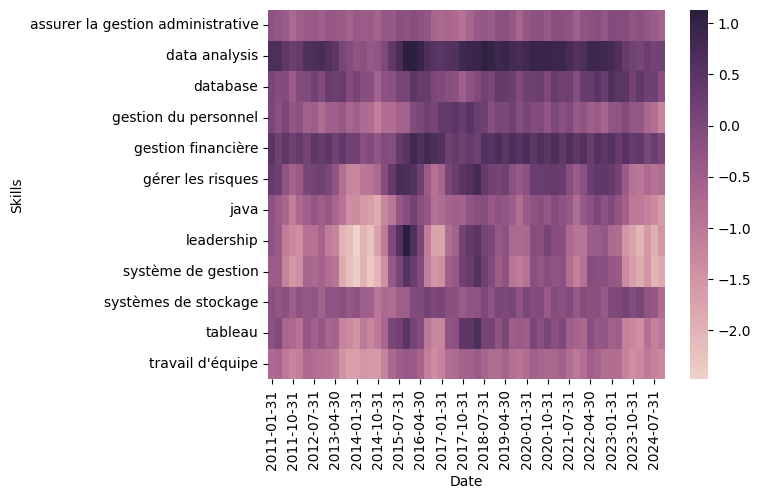}
  \caption{Heatmap showcasing the top skills for the job "Consulting Engineer in IT Management" and their level of importance through the years.}
  \label{heatmap_2}
\end{figure*}

To evaluate the inference capabilities of the model on unseen job-skill pairs we generated all possible facts between skills and specific job titles, then we removed from those facts all the facts that were seen during training. Time range was added to the resulting facts and fed to TA-DistMult to obtain factuality scores. To visualize these scores distinctively we focused on a few skills in figures \ref{da_evol}, \ref{java_evol}, \ref{heatmap_1}, \ref{heatmap_2}. 

Figures \ref{da_evol} and \ref{java_evol} showcase the temporal evolution of the skills "data analysis" and "Java" respectively over a specified time period (from January 1, 2010, to January 1, 2025). We calculate each data point by averaging all the skill scores at a certain timestamp. The plots display the temporal evolution of the average score for the "data analysis" and "Java" skills. The x-axis represents the date (timestamps generated quarterly between 2010 and 2025), and the y-axis shows the average score, which indicates the strength or relevance of the "data analysis" and "Java" skills in relation to the selected job title over time. These plots help track how these two skills evolve in importance or demand over the years, according to the model’s scores. The line shows a temporal trend, which may reveal periods of increased or decreased importance. A highlighted range, from January 1, 2023, to January 1, 2025, is framed in red. This showcases the inference of our model on unseen timestamps. We see that the skill of data analysis is highly sought after in 2023 however, it starts to see a decline in 2024. Java on the other hand continued to see a decline since 2022.

We also plot the inferred scores of skills of the job "Production Leader" throughout the years in Figure \ref{heatmap_1}. Cooler colors represent lower values of the score averages This signals a region of minimal occurrence and relevance. Warmer colours represent higher values of the score averages indicating a high-importance region of the skill at a certain timestamp.
We see that skills such as teamwork (travail d'équipe) have been consistently high from 2011 until 2024. Other skills like risk management (gérer less risques) started being less sought after in 2023. The job "Consulting Engineer in IT management" (Figure \ref{heatmap_2}) is starting to focus less on "management systems" (système de gestion) starting 31-01-2023.

\subsection{Hyperparameter analysis}
\begin{figure}[h!]
    \centering
    \includegraphics[scale=0.6]{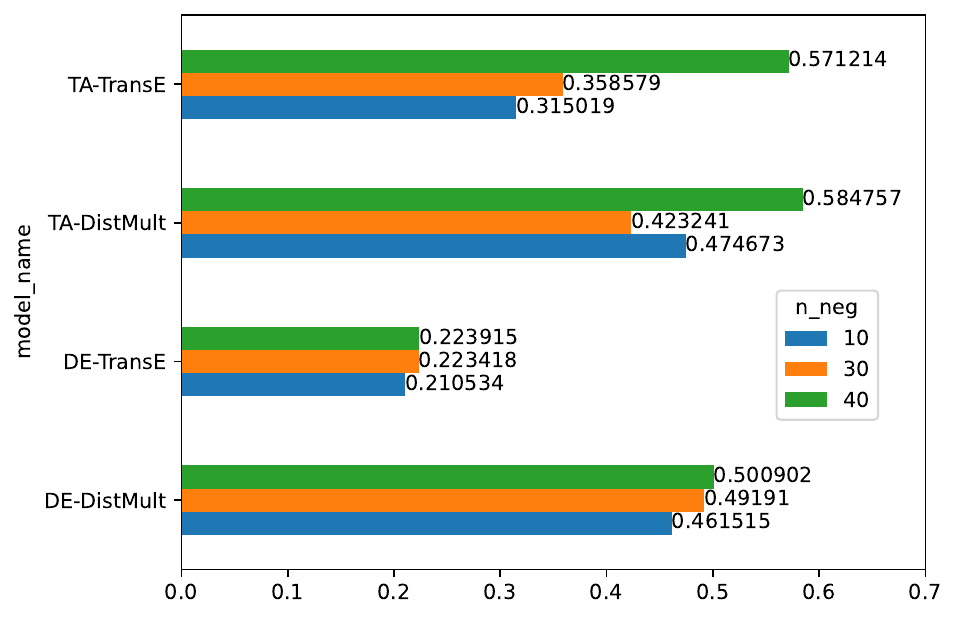}
    \caption{The average filtered MRR performance of different models on the test set using various numbers of negative examples (n\_neg).}
    \label{fig:n_neg}
\end{figure}

\begin{figure}[h!]
    \centering
    \includegraphics[scale=0.6]{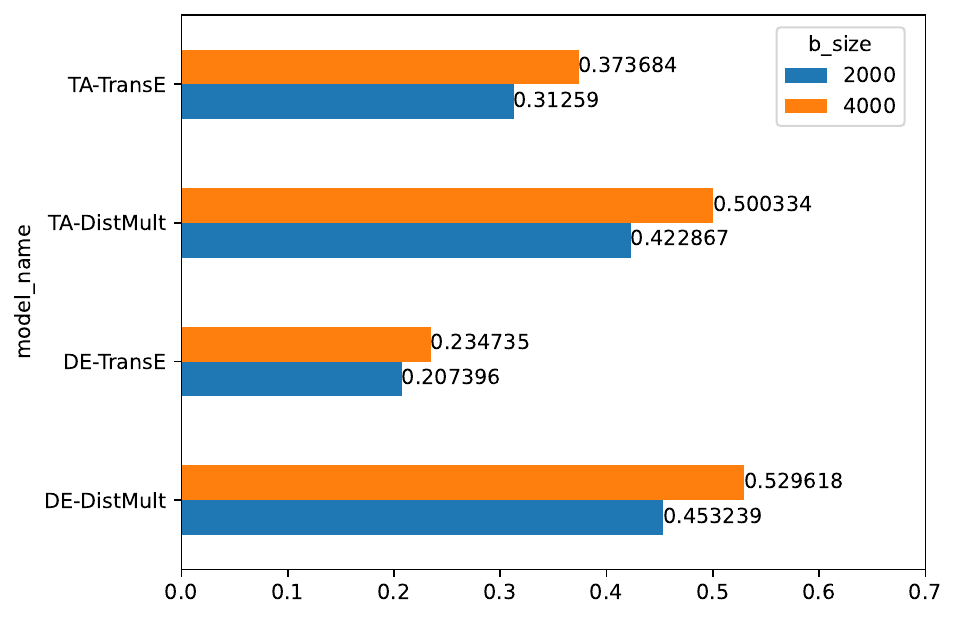}
    \caption{The average filtered MRR performance of different models on the test set using various batch sizes (b\_size).}
    \label{fig:b_size}
\end{figure}

\begin{figure}[h!]
    \centering
    \includegraphics[scale=0.6]{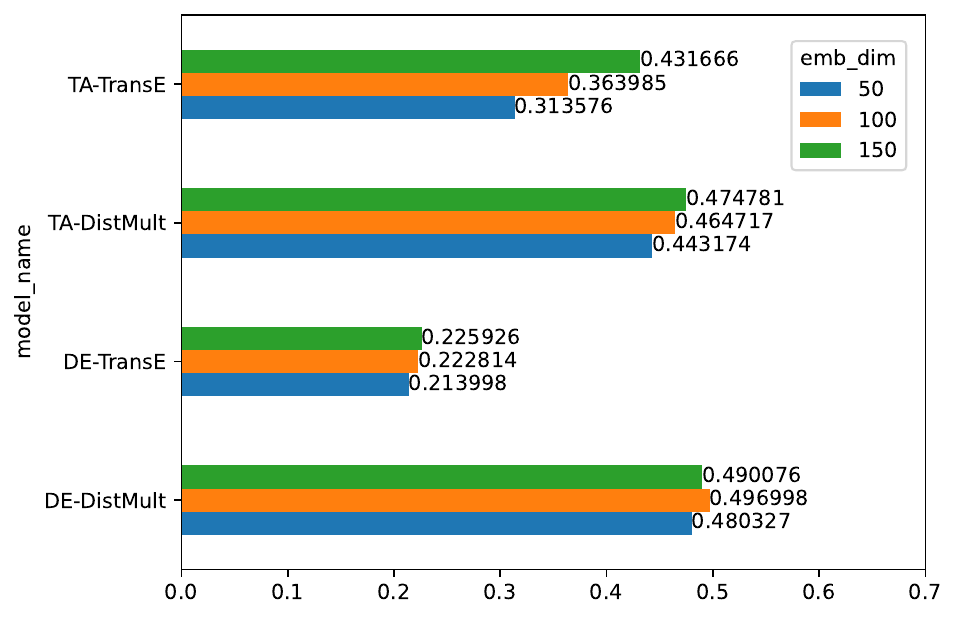}
    \caption{The average filtered MRR performance of different models on the test set using various batch sizes (emb\_dim).}
    \label{fig:emb_dim}
\end{figure}

In this section, we explore the effects of different hyperparameter settings on the different models. Figures \ref{fig:n_neg}, \ref{fig:b_size} and \ref{fig:emb_dim} show how models perform on average under different settings of negative examples, batch size and embedding dimension respectively. We observe that a higher number of negative examples generally increases the performance of the models. This can be attributed to the fact that negative examples play a role in regularisation and hence can reduce overfitting and improve generalization. Increasing the batch size also increases the performance of various models. Batch size also regularizes by counteracting the inherent variance of the models through the incorporation of more bias. We notice that increasing the embedding dimension generally increases the performance but not to the same extent as the number of negative examples and the batch size. This shows that negative sampling and batch size significantly influence model performance compared to the embedding dimension.

\section{Discussion}

In this paper, we use temporal knowledge graph embeddings to predict skill demand in the Moroccan job market. We formulate the skill need prediction problem as a link forecasting problem. We experiment with several temporal KG embedding models to achieve this task based on diachronic embeddings (DE models) and LSTM (TA models). We hypothesise that the performance of the TA models is better than the DE models due to the size and completeness of the KG. In fact, the TA approach encodes each element of the $\mathbf{r}_{seq}$ in order and feed it to an LSTM to obtain $\mathbf{r}_{seq}$ as opposed to DE which splits the embedding of entities to contain temporal and structural information. Consequently, TA has a significantly higher number of parameters which may cause overfitting especially in smaller KGs. In our case, JobEDKG is endowed with significant richness in terms of concepts, concept types and facts, which made it suitable for modelling using an overparameterized model (i.e. TA). This phenomenon is consistently measured and explored in machine learning literature and is called the bias-variance tradeoff \cite{briscoe2011conceptual}

The pipeline proposed in this paper holds potential value for organizations leveraging current skills data for implicit forecasting. Many educational institutions, including community colleges and universities, use labour market data providers like LinkedIn Talent Insights \footnote{https://business.linkedin.com/talent-solutions/talent-insights} to shape their course selections and curricula. Subscribers could gain advantages if these data providers integrate forecasting capabilities into their services.

Other organizations such as the National Center for O*NET Development, responsible for the U.S. Department of Labor-sponsored Occupational Information Network (O*NET) \footnote{https://www.onetonline.org/}, ROME \footnote{https://www.francetravail.fr/employeur/vos-recrutements/le-rome-et-les-fiches-metiers.html} and ESCO \footnote{https://esco.ec.europa.eu/en}, all of which compile occupation profiles. While these organisations already collaborate with data providers to monitor job posting data for recent changes in skill demand, forecasts demonstrated in this paper could offer early warnings before skill demand reaches a critical threshold.

Similarly, businesses operating in rapidly evolving technology sectors could find value in a skills forecasting pipeline. To maintain a relevant workforce, these companies often grapple with deciding whether to lay off existing workers and hire new ones or to invest in reskilling their current employees. Layoffs can be challenging for those losing their jobs and detrimental to the morale of remaining employees, while hiring new workers incurs significant costs. Reskilling, on the other hand, demands time, which companies often perceive as a luxury. Forecasts could provide a time buffer for reskilling, enhancing workers' well-being and improving labour market efficiency by preserving established employer-employee matches.



\section{Limitations and Future Directions}
While this study provides valuable insights into using temporal KG embeddings for skill demand forecasting, several limitations should be considered when interpreting the results. In what follows we give a list of limitations and prospective future work to mitigate them:

\begin{itemize}
    \item \textbf{Advanced scoring functions}: Our work doesn't explore other scoring functions other than TransE and DistMult. This choice was further motivated by the computational complexity of the more sophisticated approaches and the fact that the TA models gave satisfactory results that surpass what’s generally acceptable in literature by public benchmarks as a good performance. In future work, we plan on exploring more sophisticated approaches to temporal KGE to design a temporal KGE model that’s adapted to job skill forecasting.
    \item \textbf{Traditional time series}: Incorporating time series data would also be an interesting direction to explore. We intend on an extension of this approach that incorporates temporal KG embedding with traditional forecasting to solve some traditional forecasting limitations, in this case, a comparison with other traditional forecasting approaches will be conducted.
    \item \textbf{Open-world extension}: This work is the first exploration of the applicability of temporal KG embedding in job skill forecasting. Consequently, we limit our problem to a closed world knowledge graph where we focus on a fixed set of skills and jobs and how they interact semantically (e.g. is this skill that was never linked to a job useful for this job?) and temporally (e.g. how is a skill demand for a job evolve with time?). However, we intend to explore the feasibility of working in an open world knowledge graph \cite{shi2018open} where we can encounter unseen entities (jobs, skills, recruiters, sectors etc) and unseen relations among them or with seen entities during training.
    \item \textbf{Real-world integration}: 
    Integrating our model into real-world applications, such as labour market forecasting and educational curriculum design, presents several technical and operational challenges. On the technical side, issues like data interoperability, scalability, and real-time performance must be addressed to ensure the model functions effectively with diverse and large-scale data sources. Operationally, the model would need to be tailored to specific domain requirements, such as adapting to evolving labour market trends and aligning educational curricula with industry standards. Although this study focuses on validating the model within the JobEdKG dataset, future research will explore these integration challenges, aiming to optimize the model for practical deployment in dynamic environments.
\end{itemize}

\section{Conclusion}
\label{conclusion}
Skill mismatch poses a significant challenge, adversely affecting recent graduates and, by extension, the global economy. Our approach to associating skills with job titles is framed as a temporal link prediction problem. To tackle this, we employ an advanced temporal knowledge graph embedding model known for its proficiency in link prediction. Utilizing this model allows us to deduce new connections between entities within the knowledge graph and eventually mitigate the skills gap, which can help students and workers pursue their professional pathways.
\printcredits
\section*{Declaration of Competing Interest}
The authors declare that they have no known competing financial interests or personal relationships that could have appeared to influence the work reported in this paper. 

\bibliographystyle{apalike}

\bibliography{cas-refs.bib}

\end{document}